\documentclass[aip,graphicx]{revtex4-1}
\usepackage{graphics,amsmath,amssymb,graphicx}
\usepackage{xcolor}
\draft 

\begin{document}


\title{On the critical role of the rear wall thickness of a grooved TNSA target} 



\author{Imran Khan}
\email[]{imran.skd95@gmail.com}
\affiliation{Department of Physics, Indian Institute of  Technology Delhi, Hauz Khas, New Delhi, India-110016}


\author{Vikrant Saxena}
\email[]{vsaxena@physics.iitd.ac.in}
\affiliation{Department of Physics, Indian Institute of  Technology Delhi, Hauz Khas, New Delhi, India-110016}

\date{\today}

\begin{abstract}
The cutoff energy and the divergence of the protons generated by the target normal sheath acceleration mechanism are known to be significantly influenced by micrometer and nanometer-size structures on the target front and rear surfaces. Specifically, the cutoff energy is significantly enhanced by creating a central rectangular groove on the target front surface, as shown in a recent study [Physics of Plasmas, 30(6), 063102 (2023)]. Here we report on 2D Particle-In-Cell (PIC) simulations to thoroughly explore the effect of the depth of the central rectangular groove on the energy spectra of the accelerated protons. The proton cutoff energy is found to enhance drastically as the thickness of the rear wall of the groove is reduced from a few micrometers to a few tens of nanometers, however, it drops sharply as the thickness of the rear wall is further reduced towards creating a complete hole through the target.

\end{abstract}

\maketitle 

\section{Introduction}
The interaction of a high-intensity femtosecond laser pulse with a solid target results in highly energetic ions with MeV energies. These ion sources are of much interest as they offer measurement of fast-evolving electric and magnetic fields using proton radiography\cite{borghesi2002electric, borghesi2003measurement,mackinnon2006proton} technique. Other potential cutting-edge applications, in the foresight, include hadron therapy \cite{bulanov2002feasibility, ledingham2014towards}, isochoric heating of matter  \cite{patel2003isochoric}, fast ignition of fusion targets\cite{roth2001fast, atzeni2002first}, and many more.

The ion acceleration process by laser-plasma interaction unfolds as an ultra-intense laser pulse propagates through a plasma, imparting energy to charged particles and propelling them to beyond MeV energies.  Several schemes have been proposed to understand the underlying mechanisms governing the acceleration of ions in the dynamic context of laser-plasma interactions. Two extensively studied mechanisms in laser-ion acceleration, Radiation Pressure Acceleration (RPA)\cite{esirkepov2004highly, robinson2008radiation} and Target Normal Sheath Acceleration (TNSA)\cite{wilks2001energetic, snavely2000intense, mora2003plasma, passoni2008theory, passoni2010target}, are very promising but have their limitations. The radiation pressure acceleration (RPA) mechanism, which postulates the existence of a bunching field through the piling of electrons at the rear surface, faces challenges such as strong electron heating due to transverse instabilities and finite spot size. This, in turn, may induce relativistic transparency, limiting the bunching electric field and resulting in large energy spreads\cite{macchi2009light,macchi2010radiation}. On the other hand, the target normal sheath acceleration (TNSA) mechanism lacks the crucial longitudinal bunching field due to the low density of energetic electrons, leading to ion energy spectra characterized by exponential decay\cite{passoni2008theory, passoni2010target}. As the current study concerns the TNSA regime, we would focus our attention on the same.

Numerous efforts have been dedicated to overcoming the limitations of the TNSA scheme, including multi-pulse oblique incidence schemes\cite{ferri2019enhanced, khan2024not, rahman2021particle}, applying external magnetic fields\cite{arefiev2016enhanced, weichman2020generation, khan2024tnsa}, and advancements in target structures\cite{khan2023enhanced, cowan2004ultralow, schwoerer2006laser}. 
As compared to other approaches, structuring the target front and rear surfaces is a comparatively easier and less expensive approach to enhance the ion/proton energies in the TNSA regime.

Both theoretical \cite{ferri2020enhancement,klimo2011short, feng2018effects, zou2019enhancement, andreev2011efficient, zhu2022bunched, shen2021monoenergetic, sarma2022surface} and experimental \cite{mackinnon2002enhancement, wagner2016maximum, cowan2004ultralow, schwoerer2006laser, floquet2013micro, purvis2013relativistic, ceccotti2013evidence, cerchez2018enhanced, qin2022high, gaillard2011increased} investigations have been conducted in recent past to study the influence of target geometry on the properties of accelerated ions and protons. These include the role of target thickness\cite{mackinnon2002enhancement, wagner2016maximum}, effect of nanostructuring on the target rear \cite{cowan2004ultralow, schwoerer2006laser}, and diverse modifications to the target front, encompassing nanoholes\cite{ferri2020enhancement}, nanocones \cite{klimo2011short, ferri2020enhancement}, grating structures \cite{andreev2011efficient, klimo2011short}, nanowires \cite{feng2018effects, zou2019enhancement}, and nanospheres \cite{klimo2011short}. The characteristics of these structures, along with the angle at which the laser pulse interacts with them, are crucial factors that influence the proton/ion spectra. Interestingly, even at modest laser powers, these shaped targets have been demonstrated to have the upper hand on the flat targets and yield significantly higher proton/ion energies.

In the present work, we investigate the impact of the depth of a micrometer-size groove on the front side of the target, or in other words the role of the thickness of the rear wall of the grooved target, in improving proton cutoff energies and their angular divergence. In particular, we investigate the variation in proton energy spectra as the thickness of the rear wall of the groove is reduced from a few micrometers to a couple of tens of nanometers, and then to the case of no wall representing a target with a complete hole through it.
 
\begin{figure}
	\includegraphics[width=.30\textwidth]{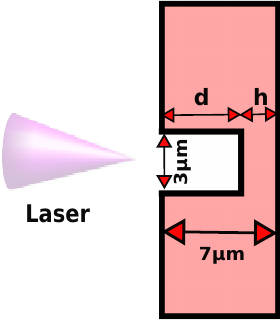}
	\caption{ Schematic representation of the target geometry used.}
   \label{fig:mimic}
\end{figure}

The manuscript is organized as follows. In the next section, the simulation details are given. The simulation results addressing the role of the groove depth on the protons' energy spectra are discussed in Section III. Finally, important conclusions of this work are presented in the last section.

\section{Simulation Setup}

In this work, we have used the fully relativistic particle-in-cell (PIC) code EPOCH\cite{arber2015contemporary} to perform Two-dimensional (2D) simulations. We select laser parameters based on the experimental studies using the GEMINI Ti:Sapphire laser at Rutherford Appleton Lab (RAL), STFC, UK, as reported in Ref.\cite{scullion2017polarization}. The laser pulse is y-polarized, with an intensity of $5.5\times 10^{20}$ W/cm $^{2}$ and wavelength of $0.8\mu$m. The focal spot size at the waist, $w_0 = 3~\mu$m, has a Gaussian profile in both space and time, and the pulse duration (FWHM) is 40 $fs$. 

In order to reduce the computation cost of these Particle-in-Cell (PIC) simulations, a fully ionized quasi-neutral polyethylene target ((C2H4)n) with protons, electrons, and C$^{+6}$ ions is considered. The mass density of the polyethylene is chosen to be $\rho = 0.93 g/cm^3 $, which translates to number densities $8 \times10^{22}$ cm$^{-3}$, $3.2 \times10^{23}$ cm$^{-3}$ \& $4 \times10^{22}$ cm$^{-3}$ corresponding to protons, electrons, and carbon ions, respectively. The number of particles per cell is chosen as 20 for carbon ions and 60 for electrons as well as protons.  

The target is localized between $\pm$ 94 $\mu$m in the transverse direction (y-axis) and between 0 to 7 $\mu$m along the longitudinal direction (x-axis) while the simulation domain is extended between  $\pm$ 94.5 $\mu$m (7000 cells) in the transverse direction and -10 to 80 $\mu$m (10000 cells) in the longitudinal direction. Open and simple laser boundaries are used on the right and left sides of the simulation box. In the transverse direction, thermal boundaries are used for particles whereas periodic boundaries are used for fields. The laser is incident normally from the left side of the simulation box.

\section{Result and Discussion}

As reported in \cite{khan2023enhanced}, in the case of a hydrocarbon TNSA target with a micron-size groove having a lateral width of 3$\mu$m (for a laser pulse having a beam waist of approximately 3$\mu$m), the cutoff energy is found to increase linearly with the groove depth. In the present work, we investigate how the spectra of accelerated protons change when the rear wall thickness (h) (shown in Fig. \ref{fig:mimic}) is reduced to a few tens of nanometers. Furthermore, the case of a target with a through hole is also studied and is compared with the case of nanometers-thick rear wall.

\begin{figure}
	\includegraphics[width=.80\textwidth]{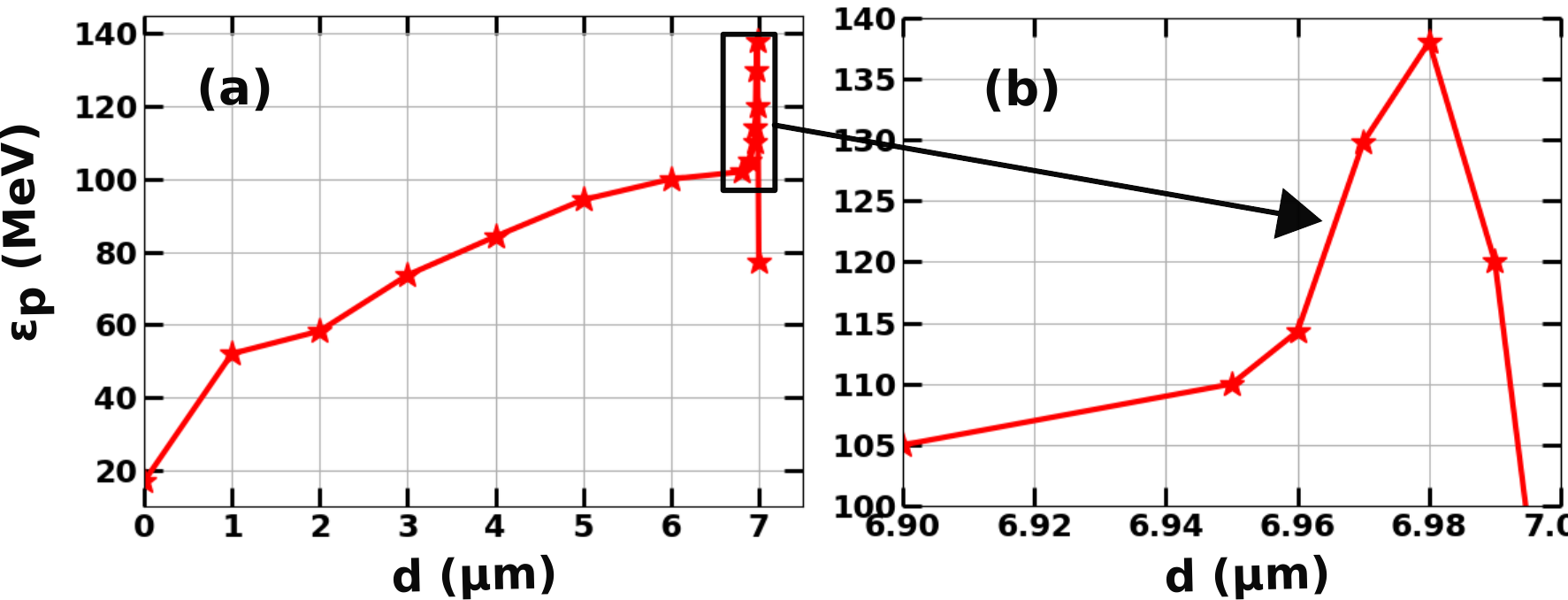}
	\caption{Proton cutoff energy with groove depth (left) and expanded view near the maximum cutoff energy (right) at t = 600fs}
   \label{fig:max_en_p}
\end{figure}

In Fig. \ref{fig:max_en_p}a, the proton cutoff energy is plotted with the increasing groove depth (or decreasing rear wall thickness). The cutoff energy increases linearly with the groove depth until d=6$\mu$m, i.e., when the thickness of the rear wall (h) is 1$\mu$m. But as the rear wall thickness is further reduced to a few tens of nanometers, the proton cutoff energy increases almost exponentially and reaches a maximum value at d= 6.98$\mu$m or h = 20$n$m. On further reducing $h$ below 20$n$m, there is a sharp decline observed in the cutoff energy. This drop in the cutoff energy continues until the rear wall is completely removed (d= 7$\mu$m, h=0$\mu$m) and there is a through hole across the target thickness. 

\begin{figure}
	\includegraphics[width=.80\textwidth]{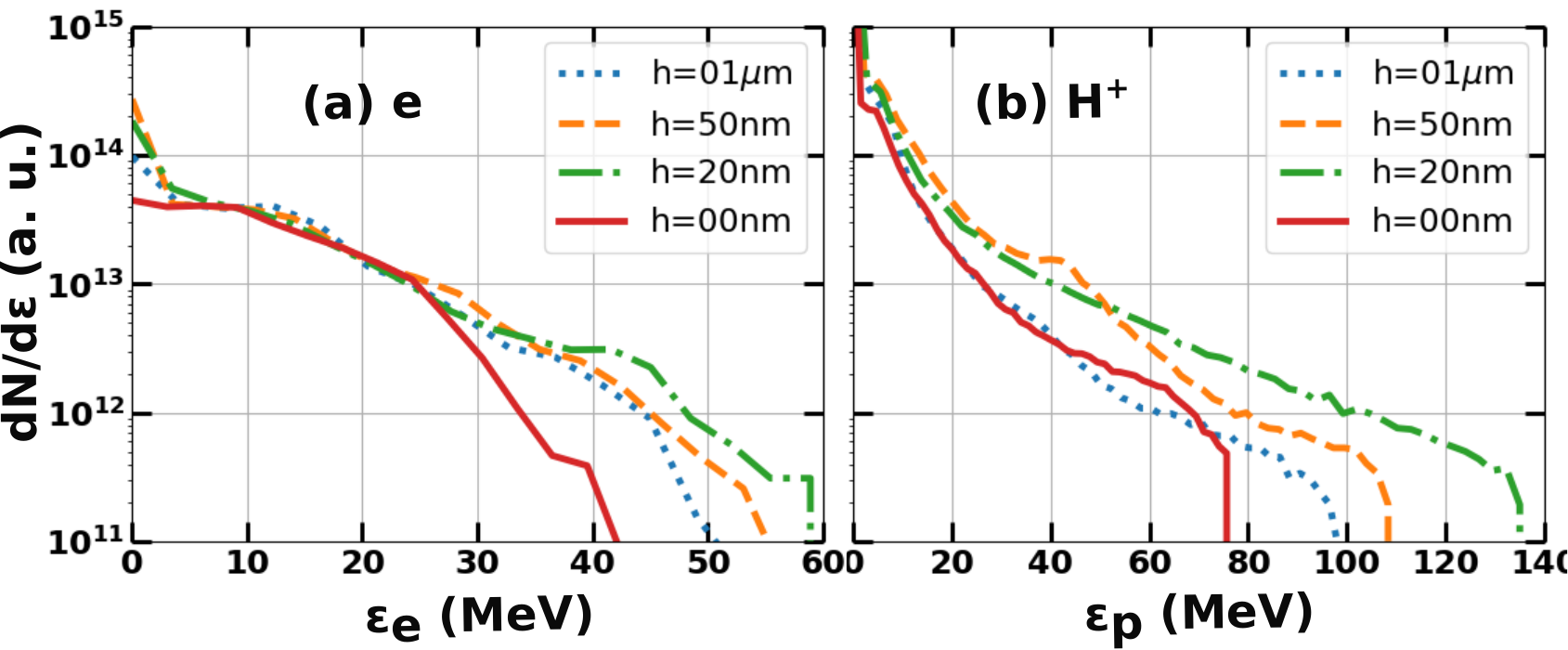}
	\caption{ Energy distribution for electron (a) at t=80fs (a), and  for proton at time 600fs (b).}
   \label{fig:en_dist}
\end{figure}

In figure (\ref{fig:en_dist}) a comparison has been shown among the electron and proton cutoff energies for the following four cases: (a) h = 01 $\mu$m (dotted line), (b) h = 50 nm (dashed line), (c) h = 20 nm (dashed dot line), and (d) h = 00 nm (continuous line) representing the complete hole case. 

 It should be noted that the electron cutoff energy continuously increases with decreasing the rear wall thickness while it is drastically reduced for the complete hole case, see Fig.\ref{fig:en_dist}a. The maximum cutoff energy is obtained for the rear wall thickness h = 20 nm, which is slightly larger than twice the collision-less skin depth (9.4 nm). A same trend is followed by the proton energy spectra, see Fig.\ref{fig:en_dist}b.  The case of h = 20nm or 50nm thick rear wall in our case is similar to the case of a hollow cone target with a few tens of nanometer foil on its apex \cite{zhu2022bunched}. In their case, the electron energy spectra are exponentially decaying, while the proton energy spectra show a sharp peak towards the high energy side. But in our case, both the electron and proton spectra exhibit exponential decay, even for the 20nm thick rear wall. The characteristic difference in the proton energy spectra may be understood by noting that in the hollow cone targets protons originate from a few nm thin slice (the cone being metallic) whereas in the present case, the protons originate from different regions in the entire hydrocarbon target.

\begin{figure}
	\includegraphics[width=.80\textwidth]{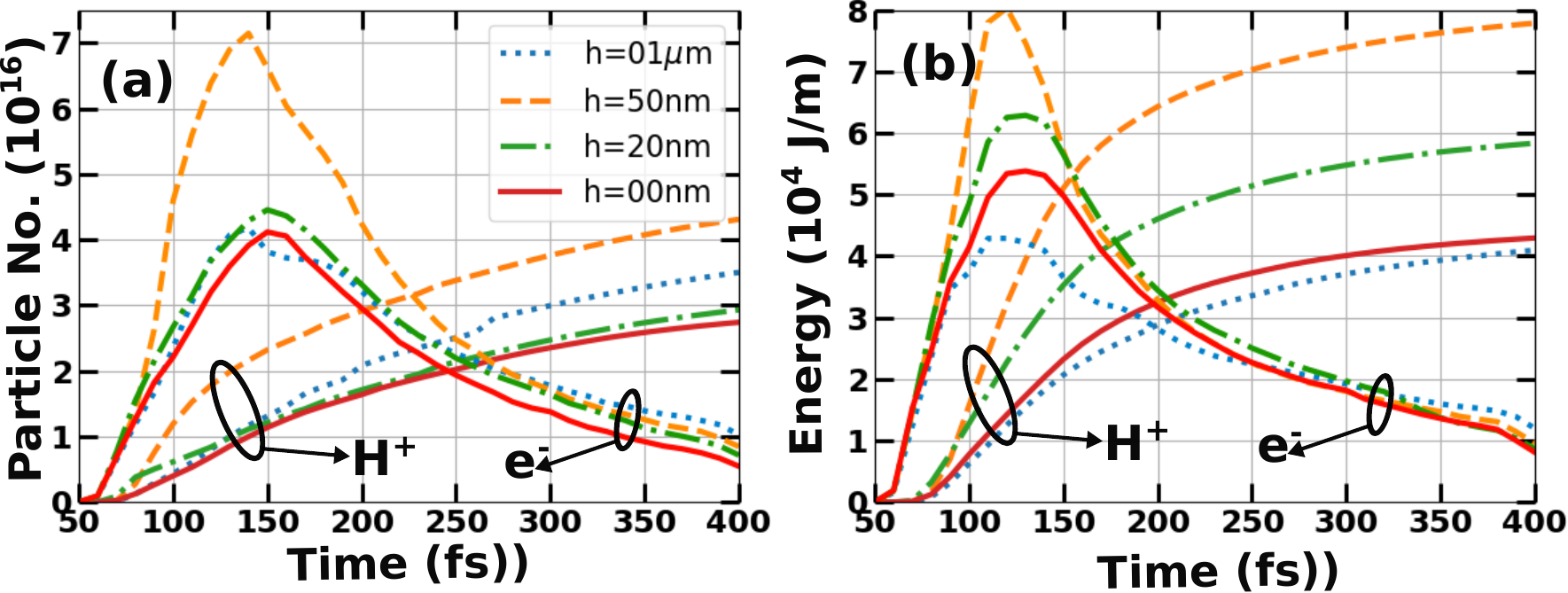}
	\caption{ The number of the electrons and protons having energy greater than 1MeV, moving toward the rear side of the target (left), and total energy carried by all electrons and protons moving to the rear side with time (right).}
   \label{fig:n_en_ep}
\end{figure}

\begin{figure}
	\includegraphics[width=1\textwidth]{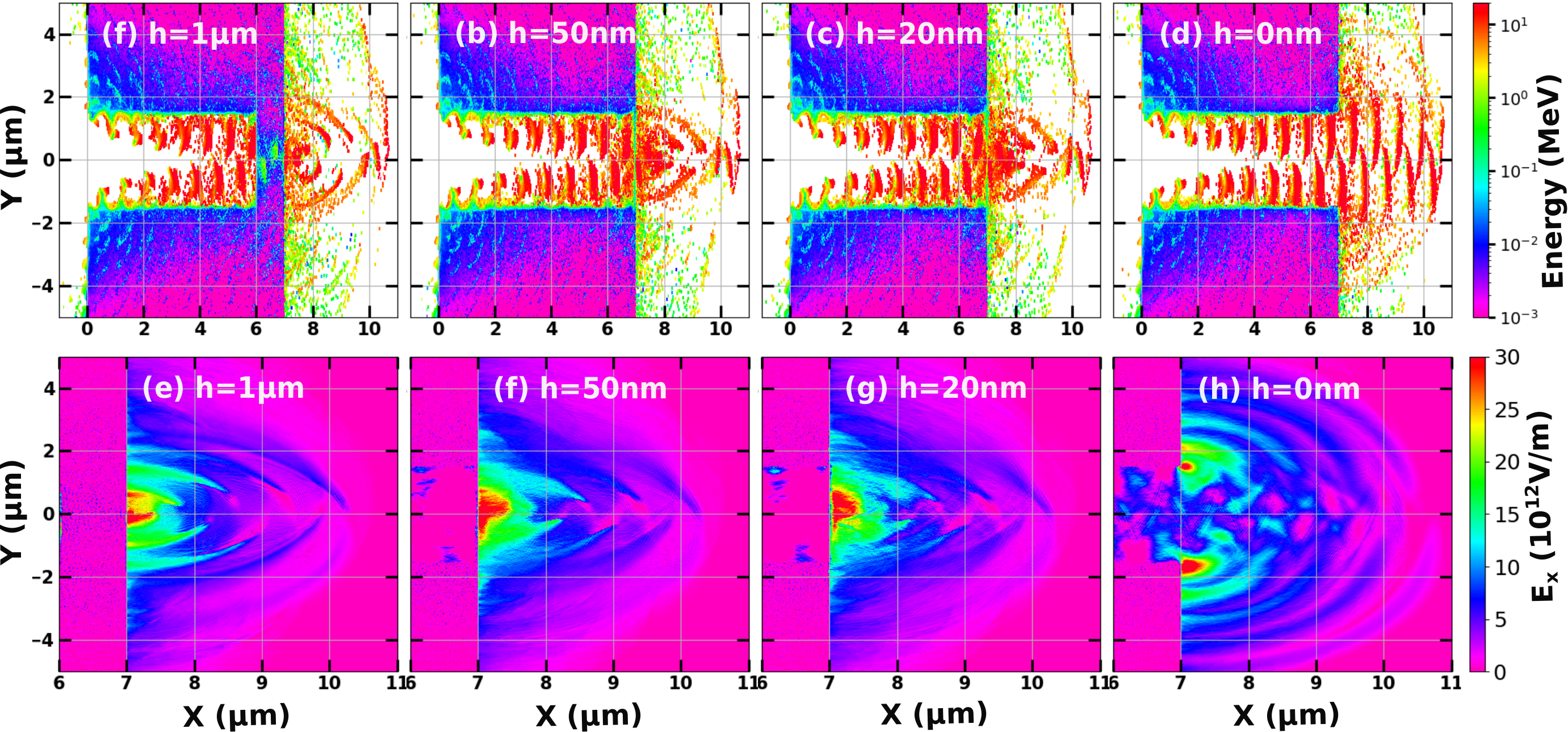}
	\caption{Average electron energy distribution in space (row one) for (a) h = 1$\mu$m, (b) h = 50 nm, (c) h = 20 nm, and (d) h = 0.0 nm at time t = 70 fs. Row two represents the corresponding rear sheath field (1e12V/m) in space at the same time.}
   \label{fig:av_en_sheath}
\end{figure}

Figure \ref{fig:n_en_ep}a shows the total number of protons and electrons (having energy above 1 MeV) moving towards the rear side of the target. The total number of electrons and protons is found to be maximum for h = 50nm. Also, the total energy carried by the electrons as well as the protons, moving at the rear side, is maximum for the h = 50nm case (Fig. \ref{fig:n_en_ep}b), but the cutoff energy for both species is maximum when h = 20nm (Fig. \ref{fig:en_dist}). 

To understand this we look at the spatial distribution of average electron energy (Fig. \ref{fig:av_en_sheath}(a-d)) and its corresponding time averaged sheath field (Fig. \ref{fig:av_en_sheath}(e-h)) at time t = 70 fs. When the laser pulse enters a grooved \cite{khan2024not,khan2023enhanced} target with an inner diameter (ID) equivalent to the beam waist size, its $E_y$ component extracts electron bunches from one side of the wall in the first half of the pulse, and from the other side in the second half. As a result, these electron bunches are separated by $\lambda/$2, coupled with a laser, and accelerated towards the rear side.

In the first three cases, h = 01$\mu$m, 50 nm, and 20 nm, the energetic electron bunches that are generated from the two side walls of the groove and move towards the rear side encounter the rear wall, which helps them focus towards the symmetry axis before exiting from the rear side of the target in hemispherical fronts which have semi-circular projections in 2D as shown in Fig. \ref{fig:av_en_sheath}(a-c). On the other hand, in the case of a complete hole (h = 0.0 nm), the energetic electron bunches generated from the two side walls of the groove freely exit the rear side where they begin to form discontinuous fronts (Fig. \ref{fig:av_en_sheath}d). So, in the first three cases, the sheath field has a well-focused configuration (Fig. \ref{fig:av_en_sheath}(e-g)) whereas in the case of a complete hole, the resultant field has a broken structure with its maxima occurring at two rear corners of the hole as shown in Fig. \ref{fig:av_en_sheath}h.

\begin{figure}
	\includegraphics[width=0.5\textwidth]{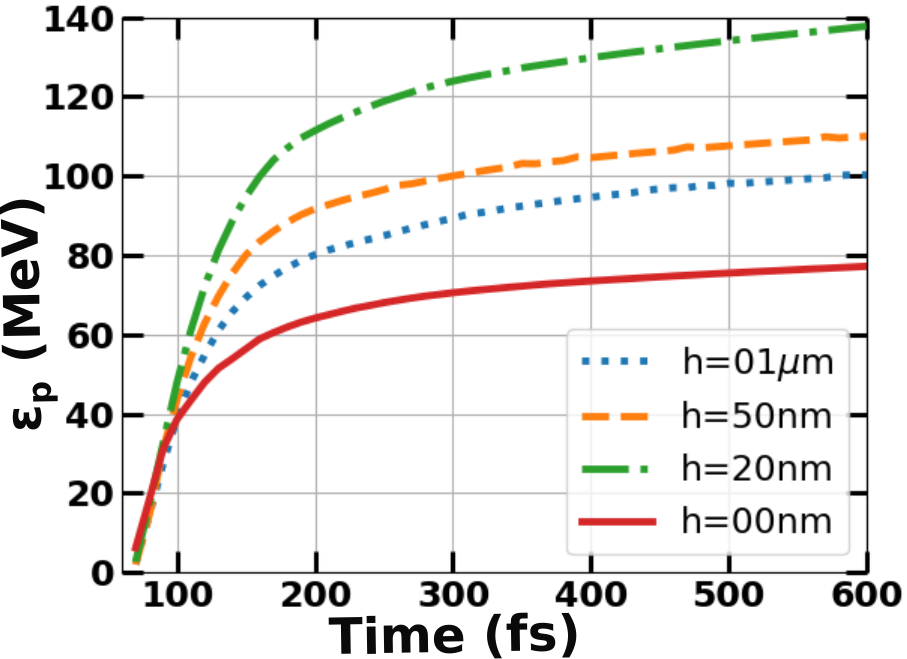}
	\caption{The temporal progression of the cutoff energy of the proton at the target's rear side.}
   \label{fig:cutoff_en_p}
\end{figure}

\begin{figure}
	\includegraphics[width=1\textwidth]{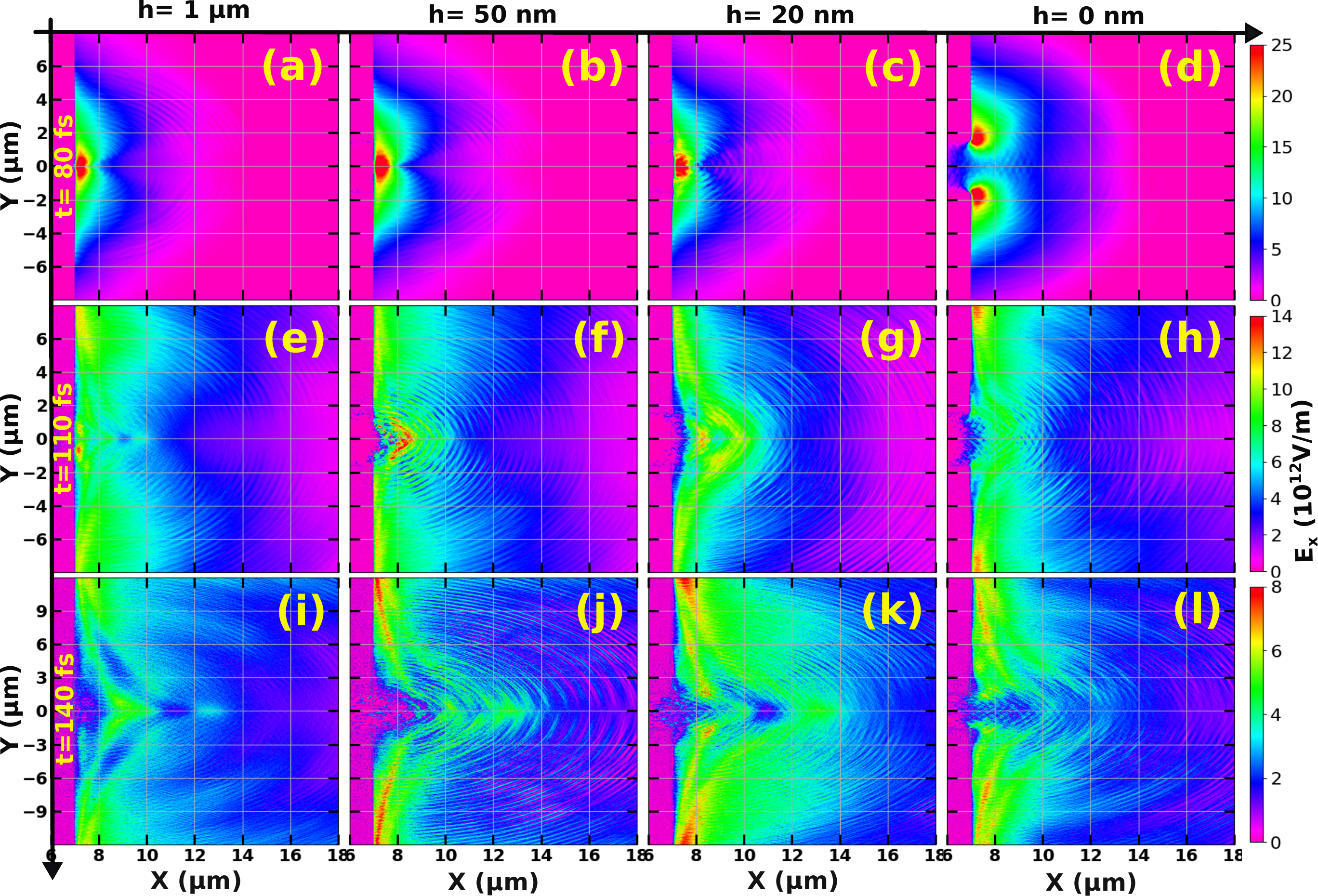}
	\caption{ The time evolution of the sheath field at the rear side of the target at time t= 80fs (a-d), t = 110 fs (e-h) and t = 140 fs (i-l)for h = 1 $\mu$m (a, e, i), h = 50 nm (b, f, j), h = 20 nm (c, g, k), and h = 0 nm (d, h, l). The sheath fields at times 80, 110, and 140 fs are the temporal average sheath fields for times 77 to 84 fs, 107 to 114 fs, and 137 to 144 fs.}
   \label{fig:sheath}
\end{figure}

Initially, up to t=80fs, the maximum value of the sheath field is almost the same (Fig. \ref{fig:av_en_sheath}(e-h) at t=70fs and \ref{fig:sheath}(a-d)) for all the cases. Therefore, the protons get almost the same cutoff energy as shown in Fig.\ref{fig:cutoff_en_p}.  At 90fs and thereafter, the cutoff energy enhancement is maximum for the h = 20 nm and minimum for the h=0nm case. This enhancement in the cutoff energy for a few tens of nanometer thick rear wall can be understood with the time evolution of the sheath field shown in Fig. \ref{fig:av_en_sheath}(e-h) and Fig. \ref{fig:sheath}. We use the temporal average of the "sheath" field over three laser cycles, as illustrated in Fig. \ref{fig:sheath}, to prevent the mixing of the actual sheath field (which accelerates the protons) with the transmitted laser fields.

The proton cutoff energy is minimum in the complete hole case (h = 0 nm) because the electrons available for re-circulation are generated only from the corners of the target. This is because there is no rear wall present in the path of the laser pulse, and the sheath field is concentrated primarily at the rear side edges of the groove/hole in the target. The electron bunches generated from the side walls overlap and create broken fronts, which is reflected in the sheath, as seen in column four of Figure \ref{fig:sheath}. Beyond this time (80 fs), the maximum value of the sheath field starts to decay. This entire scenario closely resembles the one reported by Prokopis et. al. \cite{hadjisolomou2020towards} in their study. In the absence of the rear wall/foil, the proton cutoff energy is significantly lower as compared to all other cases.

For a segment of 1 $\mu$m thickness, it significantly aids in the re-circulation of electrons, and there are no contributions from the edges of the target, which allows them to be focused towards the symmetry axis. This leads to the formation of a focused sheath field at the axis, which reaches its maximum intensity at time t = 80 fs. After that, the sheath field continuously decays as the energetic electrons start transferring their energy to protons or ions (as shown in column one of Fig. \ref{fig:sheath}). Because of the focused sheath field, protons have more energy than in the complete hole case. In both the cases (h = 1 $\mu$m and 0 $\mu$m), there is no change in the configuration of the sheath field over time; that is, the sheath field remains focused at the axis for h = 1 $\mu$m while it remains focused at the rear edges of the hole ( while developing an overlapping pattern in the axial region) for h = 0 $\mu$m.

The h = 1 $\mu$m scenario resembles the experimental results reported by Gaillard et. al. \cite{gaillard2011increased} for a hollow conical target, however, in their case, the electron bunches generated from the side walls are separated by $\lambda$ and not by $\lambda$/2. In their study, the ID of the cone (30$ \mu$m) is much wider than the beam waist size (7$ \mu$m). As a result, the laser is incident almost normally at the flattened vertex of the cone and does not interact with the side walls. This results in the cone behaving as a flat target. On the other hand, when the laser pulse is 15 $\mu$m off-axis, it interacts only with one side wall. It pulls out the electrons at alternate half cycles of the laser pulse, and hence these electron bunches are separated by $\lambda$, and not by $\lambda/2$ as in the present study. The sheath formations at the rear side of the cone target are mainly reported due to the electrons generated at the side wall of the cone\cite{gaillard2011increased,zhu2022bunched}. However, none of these reports describe the overlapped pattern in the sheath field following the breakdown of the rear wall.

\begin{figure}
	\includegraphics[width=1\textwidth]{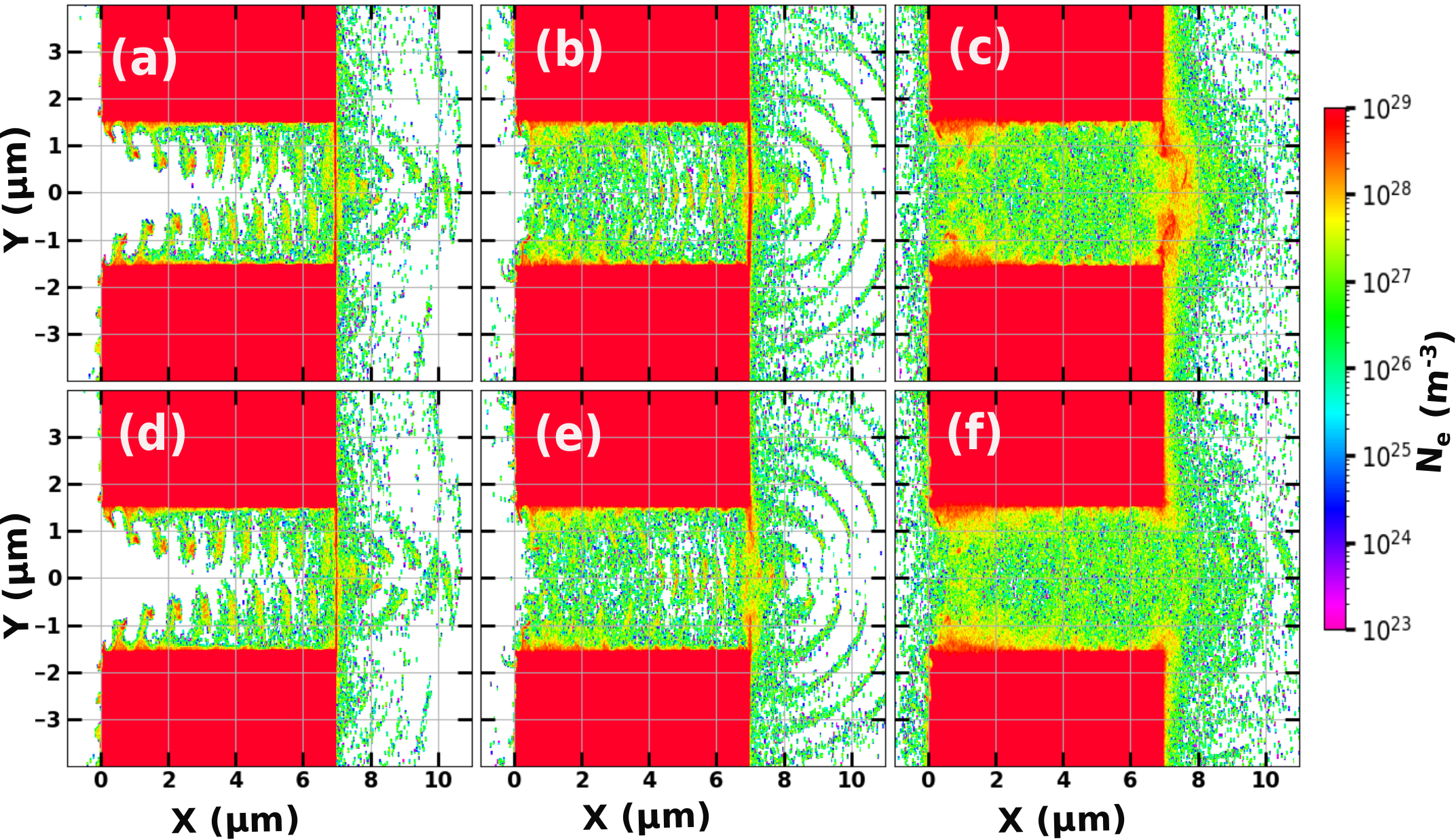}
	\caption{Distribution of electron number density in space at time t= 70fs (a, d) t = 80 fs (e, e) and t = 110 fs (c, f) for h = 50 m (a - c), and h = 20 nm (d-f).}
   \label{fig:num_den}
\end{figure}

The temporal progression of the sheath field for the nanometer-thick rear wall cases shows the transition from focused to overlapping pattern (columns 2 and 3 of Fig. \ref{fig:sheath}). The sheath field pattern for h = 20 and 50 nm starts changing its behavior from a focused (at the axis) to an overlapping pattern at time t = 80 and 110 fs, respectively. This is because the high-intensity laser pulse breaks the rear wall at these two intervals of time (t = 80 and 110 fs for h = 20 and 50 nm, respectively), as shown in Fig. \ref{fig:num_den}. The delay in time is due to the segment thickness difference (for h = 20 nm is 2.5 times thinner than that of h = 50 nm) and the Gaussian shape of the laser pulse in space and time.

As the laser pulse begins to interact with the target, its Gaussian profile causes the laser pulse front intensity to increase continuously, leading to an increase in the ponderomotive force. The h = 20 nm thick rear wall can not hold the increasing ponderomotive force for a longer time and breaks at t = 80 fs. The h = 50 nm thick wall remains for a longer time due to its 2.5 times higher thickness and breaks down at t = 110 fs.


The h = 20 nm case results in the maximum cutoff energy because initially, the focused sheath field accelerates the protons, which are then further accelerated by the overlap patterned sheath field. This overlapped sheath field in this case prevails for a longer time compared to all other cases (column 3 of Fig. \ref{fig:sheath}). This is due to the fact that the number of energetic electrons and their cutoff energy reach their maximum at h = 20 nm (Fig. \ref{fig:en_dist}a), maintaining their energy for longer time. Consequently, the sheath field is also maintained for a longer duration. 

\begin{figure}
	\includegraphics[width=1\textwidth]{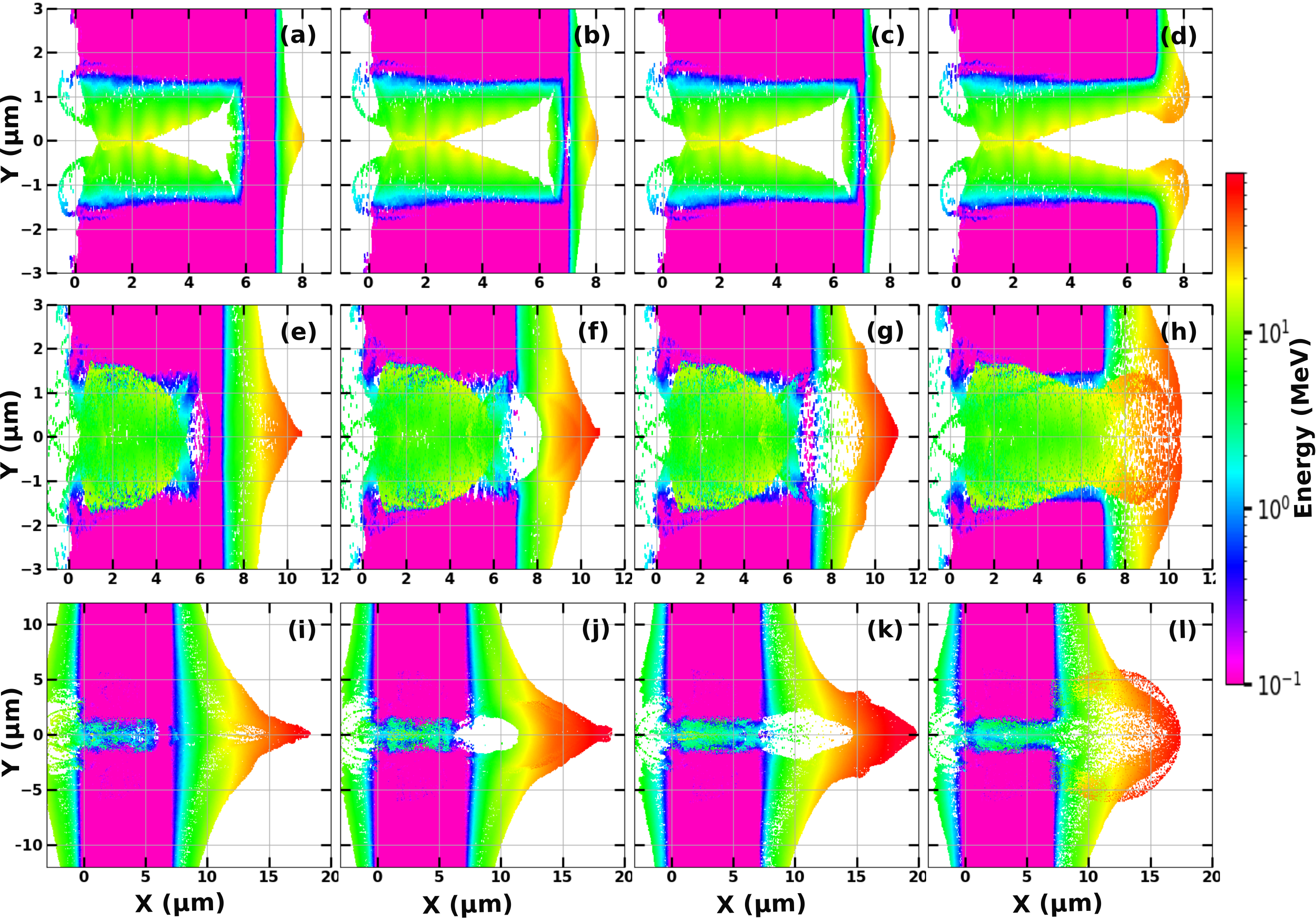}
	\caption{The time evolution of average proton energy  distribution in space at time t= 90fs (a-d), t = 120 fs (e-h) and t = 190 fs (i-l)for h = 1 $\mu$m (a, e, i), h = 50 nm (b, f, j), h = 20 nm (c, g, k), and h = 0 nm (d, h, l).}
   \label{fig:av_en_p}
\end{figure}

The h = 50 nm case results in a cutoff energy slightly lower than that for h = 20 nm. Initially, the focused sheath field accelerates the protons in the same way as that for h = 20 nm, but for a slightly longer time (up to t = 110 fs). The sheath field decays by a sufficient amount by this time, and hence the overlapped pattern that further accelerates the proton is weak and dies off more rapidly than that of h = 20 nm. Therefore, the proton's cutoff energy in this scenario is marginally lower than that of h = 20nm. Figure \ref{fig:av_en_p} illustrates the time evolution of the average energy distribution of the protons in space. At time t = 90 fs, when a segment is present in the target (Fig. \ref{fig:av_en_p}(a-c)), it seems that the energetic protons are generated from the rear wall, and their average energy distribution appears identical in all three cases. However, in the complete hole case, the sheath field is generated at the edges of the grooves (Fig. \ref{fig:sheath}d). As a result, the energetic protons originate from the edges of the groove at the rear side and begin to move in all directions (Fig. \ref{fig:av_en_p}d).

At times t = 120 and 190 fs, the evolution of proton energy for a distance of h = 1 $\mu$m follows the typical energy distribution observed in the Target Normal Sheath Acceleration (TNSA) mechanism. In the case of a complete hole, the protons originating from the corners start merging and form a very wide peak. For h = 20 nm, in addition to the main peak of proton energy, two minor peaks also appear on either side of the central peak. This is because the rear wall is very thin and gets broken in the early phase of acceleration. The sheath field is significantly strong, causing it to extract protons from the edges of the groove, which makes the beam slightly wider. However, for the thickness of 50 nm, the breaking of the rear wall occurs considerably later. At this point, the sheath field strength is significantly lower, and it is, therefore, unable to extract protons from the edges of the groove.

Finally, it is observed that the grooved target with the rear wall thickness below $20 \mu m$, the cutoff energy of protons decreases drastically, see figure\ref{fig:max_en_p}. The possible explanation of this change of behaviors may be as follows. The thickness of the target plays an important role in deciding the transparency threshold. For the target thickness, h, and normalized laser amplitude ($a_0$), $n_{th}\simeq \frac{2\lambda}{9h}(3+\sqrt{9\sqrt{6}a_0-12})n_c$ is the threshold density below which the target is transparent and above which it is opaque\cite{choudhary2016efficient}. This formula is an extension of the original formula obtained by Siminos et al.\cite{siminos2012effect} without considering the effect of the target thickness. So for the fixed laser intensity, the induced transparency would occur at a lower target density for the thick target and at a higher density for the thin target. The threshold density calculated from the local field after the propagation of the pulse inside the groove for the rear wall thickness h = 10, 20, 30, and 50 nm is 355 $n_c$, 178 $n_c$, 140 $n_c$, and 100 $n_c$, respectively. The actual density of the target is $\sim 183 n_c$, therefore the target with rear wall thickness $\geq $ 20 nm is opaque whereas for h $<$ 20 nm it is transparent to the laser field leading to a significant reduction in the efficiency of the TNSA mechanism. 

\section{Conclusion}
We have performed two-dimensional PIC simulations to examine the impact of groove depth (or thickness of the rear wall of the groove) in a target with a single, central, rectangular groove (having a lateral width equal to the laser waist size) at the front. The laser pulse extracts the electrons from the sidewalls of the groove and accelerates them in the forward direction. These electrons are accelerated for a longer time and become more focused for a higher groove depth. Therefore, the proton cutoff energy linearly increases with an increase in depth until the rear wall is of micrometer thickness.

The proton cutoff energy rapidly increases when the rear wall thickness drops to the nanometer range. The intensity of the laser pulse is high enough to break the rear wall, however, the focused sheath field accelerates the protons before the wall is broken down. As soon as the rear wall is broken down, electrons from both sides of the groove start to form an overlapping pattern in the sheath field. This leads to further acceleration of protons leading to higher cutoff energies.
However, the thin segment (h $<$ 20 nm) becomes transparent to the laser pulse even before the sheath field reaches its maximum value and therefore is not as effective for TNSA. The thick segment (h $>$ 20 nm, i.e., 50 nm), however, remains opaque, but it takes longer to break, and the overlapping pattern in the field forms at a later stage of the sheath field decay. In this case, the proton cutoff energy is significantly high but is not optimum. On the other hand, a segment of thickness h = 20 nm breaks around the time when the sheath field is at its maximum value, and an overlapping pattern is formed before the decay in its peak value leading to highly efficient proton acceleration. Therefore, a thickness of h = 20 nm is the optimal value, as it compromises between increasing the sheath field and preventing transparency. 
Finally, when there is no rear wall in the path of the laser, the sheath field is concentrated at the rear corners of the groove and is not as effective. Consequently, there is a significant reduction in the protons cutoff energy for the complete hole case.

\begin{acknowledgments}
The authors would like to acknowledge Andrea Macchi of CNR, Pisa, Italy, for his constructive comments on the manuscript. Acknowledgments are also due to the EPOCH consortium, for providing access to the EPOCH-4.9.0 framework \cite{arber2015contemporary}, and to Indian Institute of Technology Delhi for providing access to computational resources. IK also acknowledges the University Grants Commission (UGC), govt. of India, for his senior research fellowship (Grant no. 1306/(CSIR-UGC NET DEC. 2018)).  
\end{acknowledgments}
\section*{Data Availability}
The data that support the findings of this study are available from the authors upon reasonable request. 
%

\end{document}